\documentclass[prl,twocolumn,showkeys]{revtex4-1}
\pdfoutput=1
\usepackage{graphicx} 
\usepackage{amsmath,amssymb} 

\begin{document} 

\title{Local transition gradients determine the global attributes of protein energy landscapes}

\author{Francesco Rao} 
\address{Laboratoire de Chimie Biophysique/ISIS, Universite de Strasbourg, 8, alle G. Monge, Strasbourg, France}

\date{\today} 

\begin{abstract} 

The dynamical characterization of proteins  is crucial to understand  protein function. From a microscopic point of view, protein
dynamics is governed by the local atomic interactions that, in turn, trigger the
functional conformational changes. Unfortunately, the relationship between local
atomic fluctuations and global protein rearrangements is still elusive.  Here,
atomistic molecular dynamics simulations in conjunction with complex network
analysis show that fast peptide relaxations effectively build the
backbone of the global free-energy landscape, providing a connection between local and global atomic rearrangements. A minimum-spanning-tree representation, built on the base of transition gradients networks, results in a high resolution mapping of the system dynamics and thermodynamics without requiring any a priori knowledge of the relevant degrees of freedom. These results suggest the presence of a local mechanism for the high communication efficiency generally observed in complex systems.


\end{abstract} 

\keywords{Protein Dynamics, Molecular Simulations, Free-energy Landscapes, Complex Networks}

\maketitle 





In complex systems the behavior of the whole is hardly predictable from the fundamental laws of interactions of the single components \cite{anderson1972}.  Indeed, it is hard to reveal the intimate connection between the local properties of a complex system and its global behavior. This problem has been formalized in the context of complex networks \cite{boguna2008} as a question of ``navigability"; i.e., the mechanism for the efficient flow of information when the single network nodes do not have a global view of the overall topology. As a matter of fact, many collective dynamical processes are driven by the presence of (usually hidden) local gradients \cite{toroczkai2004}.


The study of protein dynamics involve a similar problem -- the coupling between the fast atomic fluctuations, which are local, and the slow conformational changes, which are global \cite{frauenfelder1991}. Those dynamical aspects have been recently recognized to be crucial for protein function, playing an important role in signalling, allosteric pathways and enzymatic reactions \cite{kern2003,eisenmesser2005}.
Molecular dynamics (MD) simulations are playing an increasing role in complementing the experimental results \cite{boehr2006,schuler2008, colletier2008} which supply useful , but limited, information to this question. The recent combination of computational and experimental studies of a protein enzyme has pointed out that the fast atomic fluctuations are partly correlated to the displacements occuring in the catalytic reaction \cite{henzler2007}. As yet, the coupling between the fast nanosecond timescales and the functional relevant transitions occuring in the microsecond to millisecond range largely remains obscure. 

Most descriptions of the free-energy surface governing protein dynamics have been rather qualitative because of the lack of
proper order parameters and the intrinsic multidimensionality of the problem \cite{du1998,pande1998}.
These limitations have triggered the development of a completely new arsenal of
tools inspired by network theory \cite{caflisch2006}. The essential idea is to map the protein
trajectory, obtained by computer simulations or experiments, on a conformation space
network (CSN), whose nodes represent the different microstates and whose links correspond to direct transitions between them \cite{rao2004,caflisch2006,gfeller2007}. This approach has been
successfully applied to the study of peptide folding and structural transitions
\cite{rao2004,krivov2004,gfeller2007,settanni2008,yang2008,prada2009}, as well as to interpret
electron transfer experiments \cite{li2008} and time-resolved IR measurements
\cite{ihalainen2007,ihalainen2008}.

In this letter, the relation between the local properties of the free-energy landscape and its global architecture is investigated by MD simulations of a 4 residues peptide, (GlySer)$_2$, and complex network analysis . In particular, when the CSN of a fully-atomistic peptide is reduced to the subgraph containing only one link per microstate pointing towards the most probable transition (i.e. following the transition gradient), the presence of energy valleys and subvalleys and their equilibrium populations is naturally extracted as well as the hierarchy of transitions between them. Hence, the fast local motions build up the backbone of the global communication. The observed coupling between local and global dynamical properties is expected to occur in a large class of complex systems.

\begin{figure} 
  \includegraphics[angle=0,width=80mm]{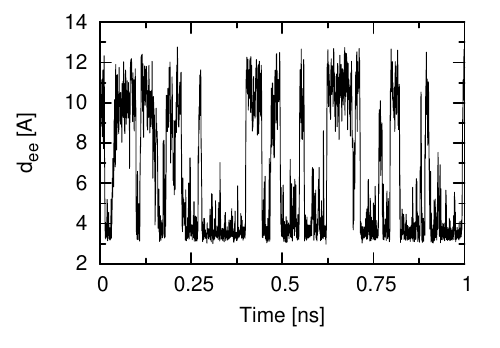} 
  \caption{Timeseries window of the end-to-end distance of the (GlySer)$_2$ peptide. The peptide is extremely flexible, alternating between compact states ($d_{ee}\sim 3.4$ \AA) and extended conformations ($d_{ee}\sim 10.8$ \AA).} 
\label{fig:timeseries} 
\end{figure}

GlySer peptides have been used for quite some time as flexible linkers (and are known to show poor secondary structure) for polypeptide dynamics \cite{bieri1999,moglich2006}. MD simulations using the Langevin algorithm 
\footnote{MD simulations, using the Langevin algorithm with a friction coefficient equal to 0.6 $ps^{-1}$, were calculated with the CHARMM program \cite{charmm2009}, the polar hydrogen energy function (PARAM19) was used. The effects of water have been included using the generalized born FACTS implicit solvation model \cite{haberthur2008}. SHAKE was employed so that an integration step of 2 fs could be used.}
and the implicit solvation FACTS \cite{haberthur2008,charmm2009} have been performed. A trajectory of 280 ns at 340 K was obtained and snapshots were saved every 140 steps for a total of  $10^6$ conformations.  During the simulation the peptide visits several different conformations characterized by an end-to-end distance between 3.1 and 12.7 \AA\ (Fig.~\ref{fig:timeseries}) indicating large structural fluctuations.
%

The peptide microstates are defined as the inherent structures (IS), i.e. the potential energy minima, of the system \cite{stillinger1982,stillinger1983}. They are calculated minimizing all the $10^6$ snapshots along the trajectory, resulting in 3044 different IS. The IS are a natural, physically meaningfull, partition into microstates \cite{baumketner2003,kim2007,rao2010}. 
The conformation space network (CSN) is built on top of this microstate definition: the nodes and the links are the microstates (i.e. the IS) and their direct transitions observed during the MD trajectory, respectively. The obtained network is weighted and it is equivalent to a classical transition matrix when the columns of the adjacency matrix (i.e. the network links) are appropriately normalized to one. 

The relation between fast local modes and global chain rearrangements is investigated by constructing from the full CSN, a new network with a reduced number of links. For each network node, the transition with the highest probability (excluding self interactions) is kept, and all others are deleted. These transitions define a gradient in the network dynamics and result in a partitioning of the network into several disconnected minimum spanning trees \cite{carmi2008}, called here gradient-clusters for convenience. (Gradient networks were originally introduced by Toroczkai and Bassler to study jamming  \cite{toroczkai2004}, though in their case the gradient is defined on the nodes as a quenched scalar field). Following the pathway defined by the most probable transitions leads to microstates lower and lower in free energy, resulting in a kind of ``steepest descent pathway" on the free-energy surface. High energy microstates, in the neighbor of a free-energy barrier, would connect either to one valley or another. As a matter of fact, the network characterizing the free-energy surface is split into a set of disconnected minimum-spanning-trees representing the local attractors of the system dynamics. 

\begin{figure} 
  \includegraphics[angle=0,width=80mm]{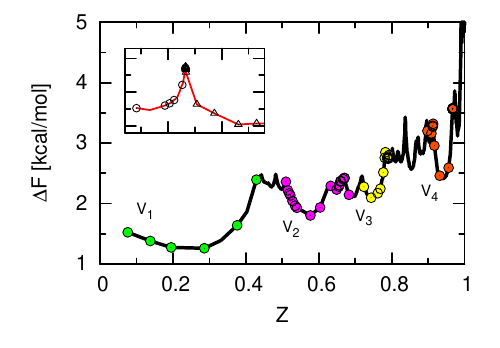} 
  \caption{(Color online) Cut-based free-energy profile for the (GlySer)$_2$ peptide. Microstates assigned by the gradient-approach to the four most populated valleys are shown on the profile in different colors. The CFEP represents the free-energy surface projected on the cumulative partition function reaction coordinate $Z$ \cite{krivov2006}, relative to a given reference microstate (in this case the most populated one).   In the inset a profile is calculated on a subgraph made by the microstates belonging to $V_4$ (triangles) and a smaller gradient-cluster (circles).} 
\label{fig:cfep-gradient} 
\end{figure}

To better understand the nature of the gradient-clusters (161 in total), a cut-based free-energy profile (CFEP) \cite{krivov2006} is calculated on the CSN and compared to the output of the gradient-partition. The CFEP is based on a network flux analysis following the idea that the network regions of minimum flow correspond to transition states \cite{krivov2006,krivov2008,muff2009}. In Fig.~\ref{fig:cfep-gradient} the calculated CFEP is shown. The profile reveals that the peptide free-energy landscape is rugged. Remarkably, the obtained gradient-clusters represent either a valley or a subvalley of the free-energy landscape (Fig.~\ref{fig:cfep-gradient}).  In Table \ref{populations} the population of the four most relevant free-energy basins detected by the gradient-approach is compared against the populations calculated by the minimum-cut method \cite{krivov2004}, which is one of the most accurate approaches for this type of calculation \cite{caflisch2006}. The results indicate that the populations of the gradient-clusters are accurate, effectively reproducing the correct thermodynamics of the system. This is particularly relevant since the gradient-approach does not use any global property of the system, neither in terms of barrier heights or microstates energies. On the other hand, the CFEP and the minimum-cut method do perform a global analysis of the network. These observations suggest an interesting application of the proposed approach to automatically detect the presence of metastable states sampled by multiple short MD runs with the aim of building simplified Markov models \cite{chodera2007}.

\begin{table} 
    \begin{tabular}{p{15mm}p{22mm}p{22mm}p{20mm}}
    \hline
    Valley    & Gradient \newline population \footnotemark[1] & Mincut\newline population  &   $<d_{ee}>$ [\AA] \\
    \hline
    $V_1$    &    0.429    &    0.428     & 4.167 \\        
    $V_2$    &    0.120    &    0.120     & 4.253 \\                
    $V_3$    &    0.068    &    0.064    & 10.960\\                 
    $V_4$    &    0.037    &    0.051    & 4.160 \\                 
    \hline
    \end{tabular}
\footnotetext[1]{To take into account of the entropic effects, gradient-clusters separated by free-energy barriers smaller than $k_BT/10$ have been merged.}  
  \caption{Comparison of the populations of the found valleys calculated by the gradient-approach or by the minimum-cut method \cite{krivov2004}. Populations are defined as the sum of the number of occurencies calculated along the trajectory of the microstates assigned to a given valley. In the last column the average value of the end-to-end distance $d_{ee}$ inside a valley is given.} 
\label{populations} 
\end{table} 

\begin{figure} 
  \includegraphics[angle=0,width=80mm]{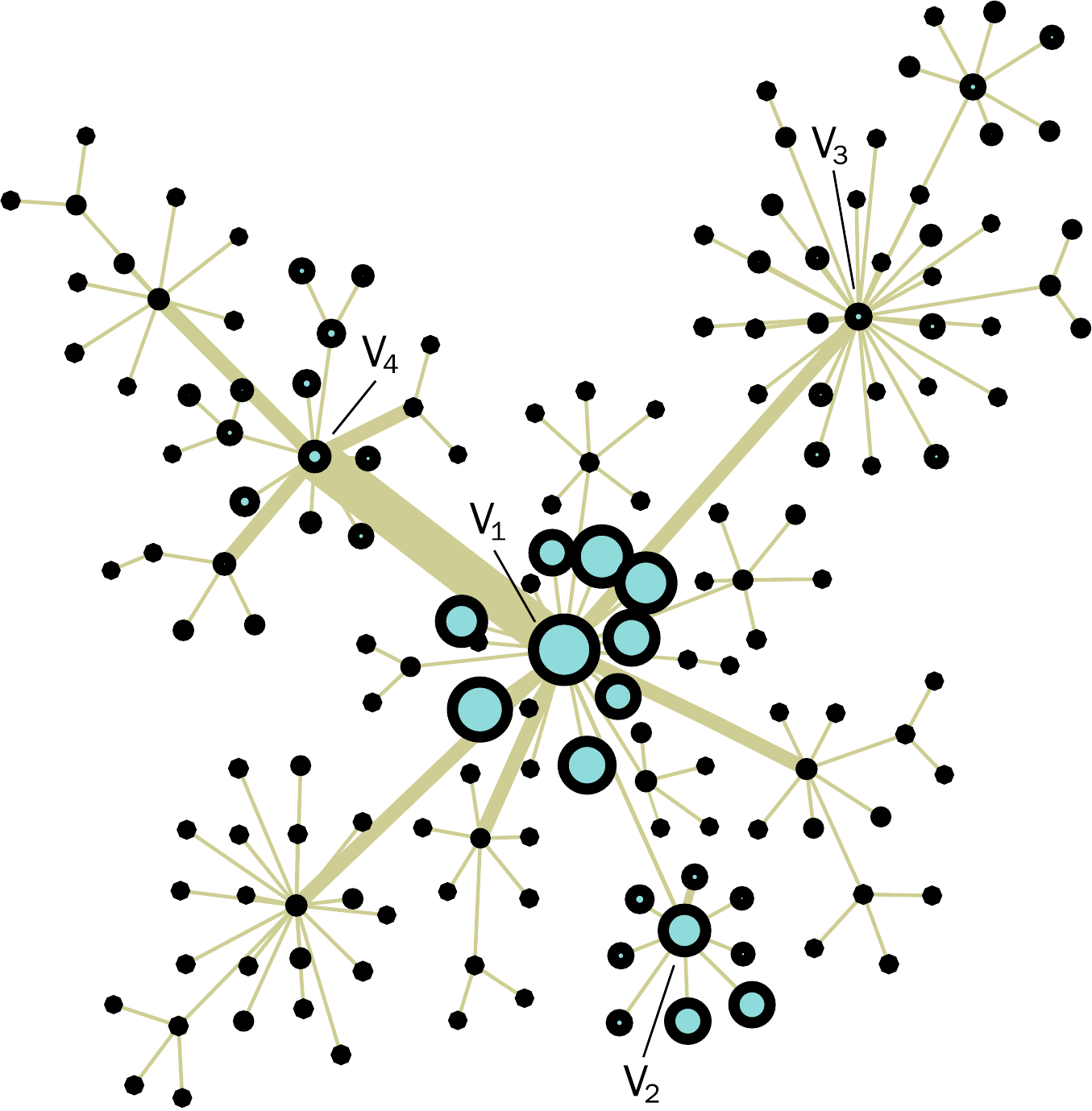} 
  \caption{(Color online) Gradient-minimum-spanning-forest of the (GlySer)$_2$ peptide energy landscape. This free-energy representation consistently represents the fast relaxations between the microstates inside a valley (star-like structures) as well as the dynamical separation between different valleys (expressed by link thickness). Larger link widths indicate slower relaxations (see main text for details). The size of the nodes is proportional to the microstate population (to avoid overcrowiding only nodes with populations larger than $10^{-3}$ are shown). } 
\label{fig:gradient-forest} 
\end{figure}

The gradient-approach can be applied in an iterative, heirarchical fashion, when the gradient-clusters are considered themselves as nodes of a higher level CSN. In this case, the nodes and the links of the network are the gradient-clusters and the connections between them, respectively. The iteration can be applied recursively until all the microstates are merged into one cluster, which is represented as a minimum-spanning-forest.  The tree structure obtained for the (GlySer)$_2$ peptide is shown in Fig.~\ref{fig:gradient-forest}. Link widths represent at which iteration step the edge was introduced. For example, the $V_1$ and $V_2$ valleys are merged at the first iteration, indicating that they interconvert rapidly. On the other hand, $V_4$ is merged to $V_1$ at the fourth iteration indicating that this is the slowest transition to $V_1$. 
The gradient-minimum-spanning-forest represents, in an intrinsically multidimensional fashion, the valleys of the free-energy landscape and their dynamical organization in fast and slow relaxations. The iteration step at which two gradient-clusters are merged together is a kinetic mesaure of the dynamical distance between the two valleys. 


Concluding, we found that, for an all-atom peptide MD simulation, the fast local relaxations produce a partition of the conformational space into disconnected minimum-spanning-trees. This partition reproduces the organization of the free-energy landscape into valleys and subvalleys with the correct populations. The iterative connection of those valleys into a minimum-spanning-forest  recovers the global backbone architecture of the dynamics occuring on the free-energy landscape. 
%
A similar coupling between local and global dynamics is expected to take place in other complex systems.
These results are relevant to investigate the still unclear relationship between network structure and dynamics in transport processes ranging from metabolic pathways to air-travelling and the internet.





\subsection*{Acknowledgments} 

I am grateful to Prof. A. Caflisch, Prof. M. Karplus,  Dr S. Muff, Dr. S. Krivov, Dr. M. Cecchini for useful comments and discussions and Prof. P. Hamm, Dr. S. Garrett-Roe for a critical reading of the manuscript.

\bibliography{/Users/ruvido/work/documents/biblio/ruvido} 

\end{document}